\documentstyle[erice99,11pt,epsfig,graphics]{article}
\input epsf

\newcommand{\be}[1]{\begin{equation} \label{(#1)}}
\newcommand{\ee}{\end{equation}}
\newcommand{\ba}[1]{\begin{eqnarray} \label{(#1)}}
\newcommand{\ea}{\end{eqnarray}}

\def\be{\begin{equation}}
\def\ee{\end{equation}}
\def\bea{\begin{eqnarray}}
\def\eea{\end{eqnarray}}

\begin{document}
\hspace{10cm}JLAB-PHY-00-03
\vspace{0.3cm}
\title{\Large Generalized Gerasimov-Drell-Hearn Sum Rule for the 
Proton-Neutron Difference in Chiral Perturbation Theory}

\author{Volker D. Burkert\footnote{e-mail: burkert@jlab.org}} 
\address{Thomas Jefferson National Accelerator Facility \\
Newport News, VA 23606} 
\maketitle

\begin{abstract}
Applications of the chiral expansion
to generalize the Gerasimov-Drell-Hearn sum rule for finite $Q^2$ are 
discussed. The
observation of several authors, that the corrections to the 
leading order contributions are large and limit the applicability 
to a very small range in $Q^2$, are only valid when considering the generalization 
of the sum rule for protons and neutrons, separately.  
When using the proton-neutron difference, the chiral expansion
may be valid to considerably higher $Q^2$ where hadronic degrees of freedom 
at large distances may connect up with quark and gluon degrees of freedom. 
This could mark the first time that nucleon structure is described by fundamental 
theory from large to small distances.
\end{abstract}

The spin structure of the nucleon has been of central interest for more than two 
decades.
Most studies have focused on the deep inelastic regime to measure the spin structure 
functions $g_1(x,Q^2)$ and $g_2(x,Q^2)$, and their respective first moments \cite{filipone}.
In recent years the interest has shifted towards the lower $Q^2$ domain and
the resonance region \cite{burli,buriof1,buriof2,softer}, 
and measurements are being untertaken 
to study the transition from the scaling regime to the regime of
 strong QCD \cite{burkert,kuhn,chen}. 
These advances in experiments made it urgent to study theoretically the connection 
between these different domains of physics. While perturbative techniques and higher twist 
expansion approaches seem appropriate at $Q^2 > 0.5 $GeV$^2$ and invariant 
masses above 
the resonance region ( $W~ > 2.5$ GeV/$c^2$), new approaches are needed to study the 
low $Q^2$ and low $W$ regions. Numerous phenomenological models have been constructed 
to describe the resonance regime and the connection with the deep inelastic regime 
\cite{burli,buriof1,buriof2,softer,ma,scholt,drech1}. 
Models that explicitely include resonances show that the resonance region, especially 
the $\Delta(1232)$, plays an important role in the helicity dependence of the 
inclusive cross section at small $Q^2$.
At $Q^2 = 0$, the sum rule by Gerasimov \cite{gerasimov}, Drell, and Hearn \cite{drell} 
(GDH SR)  relates the energy-weighted integral of the helicity-dependent cross section to the 
anomalous magnetic moment of the target nucleon:

$$ I = \int{{\sigma_{1/2}(\nu)-\sigma_{3/2}(\nu)}\over {\nu}}d\nu 
= -{2\pi^2\alpha\over M^2}\kappa^2 \eqno(1)$$ 

\noindent
where $\kappa$ is the anomalous magnetic moment of the target nucleon, 
and $M$ is the nucleon mass.  

Recently, attempts have been made to evolve this sum rule into the 
regime of finite $Q^2$ using heavy baryon chiral perturbation 
theory (HBChPT) \cite{meissner,jios}.
Ji and Osborne \cite{jios} constructed a generalized sum rule using a 
dispersion relation for the 
invariant photon-nucleon Compton amplitude $S_1(\nu,Q^2=0)$. At non-zero values 
of $Q^2$ the connection is given by the equation
$$\int G_1(\nu, Q^2){d\nu \over \nu}  = {1\over 4} \bar{S_1}(0,Q^2) \eqno(2)$$
where $G_1(Q^2,\nu)$ is the spin-dependent structure function and
$\bar{S_1}(0,Q^2)$ is the Compton amplitude, with the overline meaning that the 
elastic contribution has been subtracted. 
The right-hand side is then calculated
in HBChPT. In leading order  $\bar{S_1}(0,Q^2)$ is independent
of $Q^2$ \cite{jios}. From power counting one expects that the GDH SR 
could be evolved to a four-momentum transfer $Q^2 = 0.2$ GeV$^2$. 
Unfortunately, the next-to-leading order (NLO) calculation results in a strong $Q^2$ 
dependence \cite{jikaos} with the slope at $Q^2=0$ given by  

$${d\bar{S}_1(Q^2) \over dQ^2} = 
{g_A^2\pi\over 12(4\pi f_{\pi})^2Mm_{\pi}}[1+3\kappa_V + 
2(1+3\kappa_S)\tau^3] \eqno(3)$$

\noindent where $\kappa_V=3.706$ and $\kappa_S=-0.120$ are the experimental values of the 
isovector and isoscalar anomalous magnetic moments of the nucleon, and 
$\tau^3$ is +1 for the proton and -1 for the neutron, respectively.   
When converted to the often used dimensionless quantity 
$$\bar{I}(Q^2) = M^2\int{G_1(\nu,Q^2)} {d\nu\over \nu} \eqno(4)$$
the low $Q^2$ evolutions for the proton and neutron are given by 
\footnote{Ji et al. use $\Gamma(Q^2)$ for $\bar{I}(Q^2)$}
 
$$\bar{I}^p(Q^2) = -{\kappa_p^2 \over 4} + 6.85Q^2(GeV^2)~ +~ h.o. \eqno(5)$$ 

$$\bar{I}^n(Q^2) = -{\kappa_n^2 \over 4} + 5.54Q^2(GeV^2)~ +~ h.o. \eqno(6)$$  

The very large $Q^2$ variation in (3), (5),
and (6) is much larger than what is expected from simple power counting. 
This fact will limit the usefulness of the chiral expansion to 
very small $Q^2$ values at best. The $Q^2$ evolution predicted by Ji et al. \cite{jikaos} 
for protons and neutrons as represented in (5) and (6) is shown in 
fig. 1 compared to data from SLAC. 
For neutrons the NLO calculation predicts a sign 
change at $Q^2 \approx 0.16 GeV^2$,
pointing in the direction opposite to the high $Q^2$ data, while for the proton
the sign is correct, however with a very steep slope. If the chiral 
expansion can be applied to the individual isospin channels, it may be 
in a very 
limited  range of $Q^2 < 0.05$ GeV$^2$ only. This effectively eliminates 
the possibility of using HBChPT to connect the GDH SR for the proton and 
neutron at the photon point 
to the deep inelastic spin integrals.

It is well known that at small $Q^2$ the GDH integrals of both proton and 
neutron are dominated by the excitation of the $\Delta(1232)$ resonance.
The Delta contribution, as well as contributions of higher resonances 
are difficult to treat in HBChPT. Since the contribution of the 
$\Delta(1232)$ is very important for the individual isospin channels, we 
eliminate that contribution, as well as the ones 
of other isopin 3/2 resonance, by taking the proton-neutron difference.
This will also reduce contributions by other resonances.  
While in (3) the values in the bracket are 13.4 and 10.84 for proton 
and neutron, 
respectively, the proton-neutron difference is 2.56, yielding a five times smaller 
slope at $Q^2 = 0$ compared to the neutron and proton case. 
For the proton-neutron difference of the generalized GDH SR one obtains
 $$\bar{I}^{p-n} = {{\kappa^2_n - \kappa^2_p} \over 4} + 
1.31 Q^2 ~+~ h.o. \eqno(7)$$

In comparison with (5) and (6), a much reduced $Q^2$ dependence is 
predicted for this quantity compared to the proton and neutron, separately. 

Taking the proton-neutron difference is also quite natural in analogy with
the deep inelastic regime. While the Bjorken sum sule \cite{bjorken} for the 
proton-neutron difference has been tested experimentally \cite{review} 
with good accuracy, the corresponding Ellis-Jaffe sum rule for proton and 
neutron separately \cite{elljaf} is significantly violated \cite{review}. 
Only the Bjorken sum rule may thus be used  
to provide a reliable theoretical constraint in the deep inelastic region. 

In order to compare with existing data we convert (7) to the usual 
first moment $$ \Gamma^{p-n}_1(Q^2) = {Q^2\over 2M^2} \bar{I}^{p-n} \eqno(8)$$ 
In fig. 2, $\Gamma^{p-n}_1(Q^2)$ is shown with the data from SLAC and the
pQCD evolution of the Bjorken sum rule to order $\alpha_s^3$ \cite{slac}. 
The NLO term in the chiral expansion for the proton-neutron 
difference has now the correct sign and reproduces better the trend of the data.

A similar conclusion can be drawn for the NLO 
expansion by Ji et al. \cite{jikaos} as 
applied to the generalized GDH sum rule proposed by Bernard et al. 
\cite{meissner}. A much reduced $Q^2$ dependence is obtained for the proton-neutron 
difference in this case as well.

In order to better understand the convergence of the chiral expansion at finite 
$Q^2$, 
it is essential to evaluate the next-to-next-to-leading order (NNLO) corrections 
for the proton-neutron difference. 
With the accurate data expected at small and medium $Q^2$ from experiments 
at Jefferson Lab, stringent tests of these predictions will be possible. 
From the higher $Q^2$ end, the higher twist operator product expansion of 
pQCD may be used to extend the range down to possibly
$Q^2 = 0.5$ GeV$^2$ \cite{jikaos}. If the remaining gap between the chiral expansion 
and the higher twist expansion of QCD can be bridged  
using QCD lattice calculations
it would mark the first time that nucleon structure is described within 
fundamental theory from small to large distances, a worthwhile goal.  

In conclusion, the chiral expansion may be used sucessfully to 
expand the generalized Gerasimov-Drell-Hearn Sum Rule to finite $Q^2$
if the proton-neutron difference is used rather than  
proton and neutron separately. In the latter cases 
the NLO terms are sufficiently large to limit 
the expansion to $Q^2 < 0.05$ GeV$^2$, while in the former case 
a five times larger $Q^2$ range is obtained. It is essential to 
calculate the NNLO terms to see whether this trend is continued in 
higher orders of the chiral expansion. 

\section*{Acknowledgments}

I gratefully acknowledge encouraging comments by Xiangdong Ji and Ulf Meissner 
regarding this report.

\vspace{0.7cm}

\noindent{Figure 1.}

\vspace{0.3cm}

\noindent{First moments $\Gamma_1(Q^2)$ of the polarized structure 
functions $g_{1p}(x,Q^2)$ and $g_{1n}(x,Q^2)$. The data points are from 
SLAC, circles - proton, squares - neutron. The lines at high $Q^2$ 
are pQCD evolutions of the asymptotic behavior to O($\alpha_s^3$). 
The curves labeled ``proton'' and 
``neutron'' are predictions of the chiral expansion in 
next-to-leading order (see text). The arrow labeled ``GDH'' represents the slope of 
$\Gamma_1p$ at $Q^2=0$ as predicted by the GDH sum rule for protons.}      

\vspace{1cm}

\noindent{Figure 2.}

\vspace{0.3cm}

\noindent{{First moment difference $\Gamma^p_1 - \Gamma^n_1$. 
Data are from SLAC. Solid line labeled ``Bjorken'' represents the Bjorken sum rule, 
corrected to O$(\alpha^3_s)$ \cite{review}, the line labeled ``ChPT'' represents eqn.(7)
and (8) for the proton-neutron difference of the chiral expansion, and
the arrow represents the slope defined by the GDH SR at the photon point.}

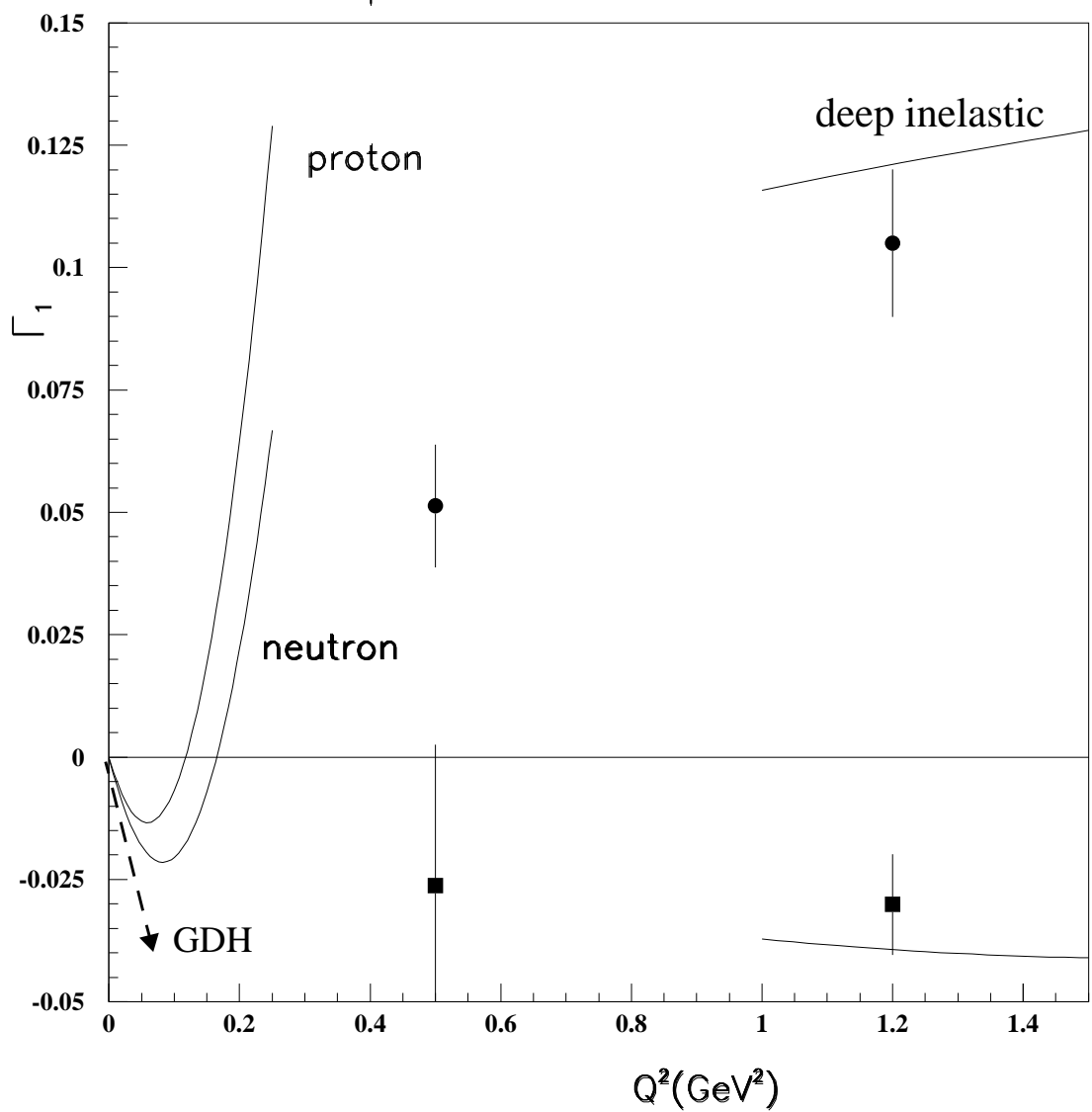
\end{document}